\documentclass[fleqn,twoside]{article}
\usepackage{espcrc2}

\usepackage{graphicx}
\usepackage[figuresright]{rotating}

\newcommand{\AmS}{{\protect\the\textfont2
  A\kern-.1667em\lower.5ex\hbox{M}\kern-.125emS}}

\usepackage{graphicx}
\usepackage{epsfig,rotating}
\usepackage[dvips]{color}

\def\be{\begin{equation}}
\def\ee{\end{equation}}
\def\ba{\begin{eqnarray}}
\def\ea{\end{eqnarray}}

\def\lsim{\raise0.3ex\hbox{$\;<$\kern-0.75em\raise-1.1ex\hbox{$\sim\;$}}}
\def\gsim{\raise0.3ex\hbox{$\;>$\kern-0.75em\raise-1.1ex\hbox{$\sim\;$}}}

\def\theta{\vartheta}

\def\dCG{{A}_{\rm CCG}}
\def\d{\rm d}
\def\bu{{\mathbf u}}
\def\bp{{\mathbf p}}
\def\br{{\mathbf r}}
\def\bpp{{\mathbf p}^\prime}
\def\brp{{\mathbf r}^\prime}
\def\pp{p^\prime}
\def\fp{f^\prime}
\def\Ep{E^\prime}
\def\Ip{I^\prime}

\def\ap{\approx}

\title{Anisotropies and clustering of extragalactic cosmic rays}

\author{M.~Kachelrie\ss
\address{Institutt for fysikk, NTNU, N--7491 Trondheim, Norway} 
}

\date{\today}

\begin{document}

\begin{abstract}
Deviations from isotropy have been a key tool to identify the sources
and the primary type of cosmic rays (CRs) at low energies. We argue that
anisotropies due to blind regions induced by the Galactic magnetic field,
the cosmological Compton-Getting effect, medium-scale anisotropies
reflecting the large-scale distribution of CR sources and the small-scale
clustering of the CR arrival directions at the highest energies
may play the same role for extragalactic CRs.
\vspace{1pc}
\end{abstract}

\maketitle

\section{Introduction}
Ultrahigh energy cosmic ray (UHECR) physics has gained increasing
momentum both on the experimental and the theoretical side~\cite{reviews}.
In the latter area, the attention has shifted from ideas involving
new physics as explanation for the AGASA excess to the study of
signatures of extragalactic cosmic rays and attempts to identify their 
sources. An important example of this development is the
reinterpretation of the ankle as dip 
produced by $e^+e^-$ pair production of extragalactic UHE protons on
CMB photons suggested in Ref.~\cite{dip}. As a consequence, the
transition from Galactic to extragalactic CRs should occur at much
lower energies than previously thought, offering experiments the
possibility to study CRs from cosmological distances with high
statistics. 

The results of the two experiments with the largest exposure until
2005, the ground-array AGASA and the fluorescence experiment HiRes, point in
several respects (GZK suppression, small-scale clusters, correlations,
anisotropy towards the Galactic center) into different
directions~\cite{x}. These discrepancies are in most cases
statistically not very significant because of the relatively small
number of events, but they obstruct a consistent interpretation of
the data. While the present state of observations is thus still puzzling,
new experiments like the Pierre Auger Observatory (PAO)~\cite{PAO},
the Telescope Array (TA)~\cite{TA} and the JEM-EUSO project~\cite{EUSO}
are expected to shed light on some of
the unresolved issues with their improved detection techniques and
increased statistics. 

Reducing experimental uncertainties and increasing the number of observed
events is obviously a pre-requisite to unravel the conundrum of
UHECRs. However, for many questions the use of the optimal tools to
analyze the data may be crucial. An example is the transition between
galactic and extragalactic CRs that is usually searched for by 
studying  the chemical composition of CRs. Here, one assumes that
the end of the galactic CR spectrum is heavy (motivated by
confinement and acceleration arguments) and that the extragalactic
component is light. At present, uncertainties in the
hadronic interaction models obstruct a reliable differentiation
even between the extreme cases of proton and iron primaries at
energies of $10^{18}$~eV and higher, as it is witnessed by the
differing conclusions in Ref.~\cite{chemie}. 
Even worse, this method fails completely if the
extragalactic component is also dominated by heavy nuclei.  

A complementary tool to study the transition from galactic to
extragalactic cosmic rays is the cosmological Compton-Getting effect (CCG)
suggested recently in Ref.~\cite{CGG}. This anisotropy is not
only a signature for CRs originating from cosmological distances but
can serve also a tool to  determine the mean charge of UHECRs
primaries and the galactic magnetic field, as will be discussed in Sec.~3 
of this short review.  

The CCG effect requires that inhomogeneities in
the source distribution of CRs are averaged out. As the free mean
path of CRs decreases for increasing energy, anisotropies connected to 
the large-scale structure (LSS) of CR sources should become more prominent
and replace the CCG. Reference~\cite{msc} analyzed the available data set
of CR arrival directions and found evidence for anisotropies on the scale of
30~degrees, consistent with the theoretical expectations for
anisotropies associated with the LSS from Ref.~\cite{napoli}.  These
results are discussed in Sec.~4. 

Finally, at sufficiently high energies deflections in magnetic fields
become negligible and a small number of bright point sources results in 
small-scale clusters of arrival directions around or near the true
source positions. Accumulating enough events, the identification of
sources will become possible using e.g.\ correlation studies. Various
studies have been pursued in this direction~\cite{ssca,sscb,ssccd,ssce,sscf,corra,corrb,corrc,corrd,corre} and are
briefly reviewed in Sec.~5. 
We start this short review of possible anisotropies of extragalactic
CRs with a discussion of the role of galactic and extragalactic
magnetic fields in the next section.

\section{The role of galactic and extragalactic magnetic fields}

The {\em galactic magnetic field} (GMF) consists of a regular and a
stochastic  component. The latter averages out along the trajectory of 
a CR and affects the arrival direction of most CRs only mildly.
Also the impact of magnetic lensing~\cite{lens} is alleviated by the 
energy dependent magnification and position of caustics, if an event 
sample cannot be binned sufficiently fine in energy.

A generalized version of the Liouville theorem  for CRs propagating in
magnetic fields  
ensures the constancy of the phase space volume along the particle
trajectories: When the density of CR trajectories is increased by
the GMF, the angular spread of their velocities increases also, so
that the CRs arrive from a larger solid angle. Both effects
compensate each other in the flux per unit solid angle, and as a
consequence an isotropic flux remains isotropic to an observer
behind a magnetized environment. However, the GMF introduces
anisotropies for an isotropic flux outside the Galaxy, if  blind
regions on the external sky exist for an observer.

A simple analytic estimate of this effect can be given e.g. for a dipole
field~\cite{kst}. Because of the azimuthal symmetry, the St{\o}rmer theory 
can be applied to determine the rigidity cutoff ${\mathcal R}_{S}$ below
which no particle can reach the Earth from a certain direction. 
Since the Earth is at zero
galacto-magnetic latitude, one obtains ${\mathcal R }_{S}=
(\epsilon\mu_G)/(2R_0^2)$, with $\epsilon\leq 1$ depending on the
arrival direction of the CR. If the
tiny vertical component detected at the solar system of $0.2\,\mu$G is
due to a dipole field, then $\mu_G\simeq 120\,\mu$G and 
${\mathcal R}_{S}$ varies in the range $10^{17}\,$V--$10^{18}\,$V. 
Although the geometry of the GMF is more complicated than a simple
dipole, one expects qualitatively similar results for more realistic
models of the GMF. Comparing the Larmor radius to the thickness of the
Galactic magnetic disk, ${\mathcal O}(100\,{\rm pc})$, shows that for
$B\simeq$ few $\mu$G particles with ${\mathcal R}\lsim 10^{17}$~V are
likely to be trapped. Numerical calculations confirm this
estimate~\cite{kst}, although precise quantitative statements depend
on the GMF considered. Note that the argument can be turned
around: For a given rigidity cutoff ${\mathcal R }_{S}$, large-scale
anisotropies should be seen around $E\sim Ze\,{\mathcal R }_{S}$, if
an extragalactic component dominates at this energy. Thus, models
that invoke a dominating extragalactic proton component already at
$E\simeq 4\times 10^{17}$~eV or extragalactic iron nuclei at $E\lsim
10^{19}$~eV might be 
inconsistent with the observed isotropy of the CR flux.

\begin{figure}
\hskip-1.5cm
\epsfig{file=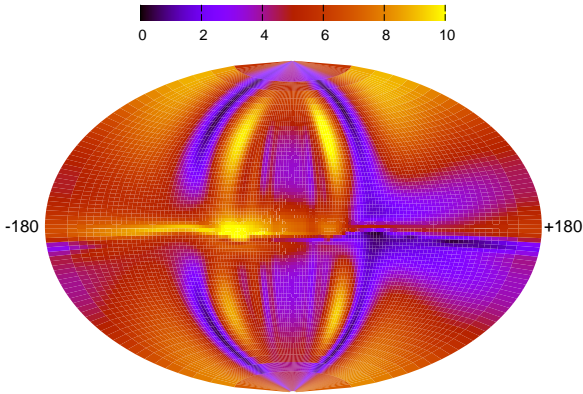,width=1.35\columnwidth} 
\vskip-1.4cm
\caption{Deflection map  of the GMF for a rigidity
  of 4$\times$ 10$^{19}$~V (for details of the model see \cite{kst,HMR}). 
  The deflection scale is in degree, the map
  refers to the direction as observed at the Earth and uses a
  Hammer-Aitoff projection of galactic coordinates.  
\label{fig:defl}}
\end{figure}

Searches for point sources of charged CRs are affected by deflections
in the GMF.  Figure~\ref{fig:defl} shows a deflection map from
Ref.~\cite{kst} for a specific GMF model presented in~\cite{HMR}
and a rigidity of $4\times 10^{19}$~V.  
The map refers to the direction as observed at the Earth and uses a
Hammer-Aitoff projection of galactic coordinates.
The expected deflections depend strongly on the direction, but
typically trajectories passing the Galactic center or plane suffer
larger deflections. Moreover, the magnitude as well as the direction
of the deflections depend on the GMF model considered.
Thus an experiment on the southern hemisphere like
the PAO might optimize  searches for point sources by cutting out
part of its field of view---or might try to correct for deflections in
the GMF and thereby test specific models for the GMF.

The trajectory of a charged CR in the {\em extragalactic magnetic
field\/} (EGMF) has the character of a random walk, each step
approximately an arc of curvature radius equal to  the Larmor radius
and size  equal to the correlation length  $L_{c}$ of the corresponding patch
of the EGMF. Therefore the average deflection is zero and the
root-mean-square deflection $\delta_{\rm rms}$ is 
$$
 \delta_{\rm rms} \ap
  0.2^{\circ}\;\frac{4\times 10^{19}{\rm eV}}{E} \,
  \frac{ZB_{\rm rms}}{10^{-11}{\rm G}}
  \left( \frac{L}{\rm Gpc}\, \frac{L_c}{{\rm Mpc}}\right)^{1/2} \,.
$$

For a calculation of $\delta_{\rm rms}$ one needs either
observational data or theoretical predictions both for the magnitude
and the structure of EGMFs. Observational evidence for EGMFs has been
found only in a few galaxy clusters observing 
their synchrotron radiation halos or performing Faraday rotation
measurements. The two methods give somewhat different results for the
field strength in clusters, with $B\sim 0.1$--1$\:\mu$G and $B\sim
1$--10$\:\mu$G, respectively. Outside of clusters only upper limits
exist for the EGMF. 

A successful predictions of EGMFs requires a
convincing theory for its origin and its amplification mechanism, but
such a theory has not emerged yet.
The seed fields of EGMF could be created in the early
universe, e.g. during phase transitions, and then amplified by MHD
processes. Alternatively, an early population of starburst galaxies or
AGN could have generated the seeds of the EGMFs at redshift between five
and six, before galaxy clusters formed as gravitationally bound
systems. In both cases, a large fraction of the universe may be filled
with seed fields for EGMFs. A quite different possibility is that the ejecta of
AGN magnetized the intra-cluster medium only at low redshifts, and
that thus the EGMF are confined within galaxy clusters and groups. 
Other mechanisms have been suggested and hence no unique model with
unique predictions for the EGMF exists.

In the last few years, magnetic fields have been included in 
simulations of large scale structures~\cite{ems,dgst,iub}. These
simulations differ both in the input physics (seed fields, 
amplification mechanism), the numerical algorithms and the extraction
of the results. Therefore, it should be not too surprising that
their result disagree strongly: While charged particle
astronomy may not be possible according to Ref.~\cite{ems}, the
deflections found in Ref.~\cite{dgst} are small in a large part of the
sky.   

Instead of viewing the results of these simulations as predictions for
the EGMF, it is more appropriate to see them as tests if the used
origin of the seeds field and amplification mechanism can reproduce
the observational data. It is more likely that future UHECR data
might teach us something about the EGMF and its origin than vice
versa. Another way to detect in particular extremely small magnetic
fields in voids is the search of extended TeV gamma-ray sources
suggested in Ref.~\cite{Neronov:2006hc}.

\section{Cosmological Compton-Getting effect}
\label{ccg}

Compton and Getting first discussed that a relative motion of observer
and CR source results in an anisotropic CR flux, using this effect
as signature for the Galactic origin of CRs with $E\gsim 0.1\:$GeV.
Similarly, the movement of the Sun relative to the microwave
background frame induces a dipole anisotropy in any diffuse cosmic
flux. 

Lorentz invariance requires that the phase space distribution function
$f$ in the frame of the observer, $f'(\brp,\bpp)$, equals the one in
the frame in which the UHECR flux is isotropic,
$f(\br,\bp)$. Expanding in the small parameter $\bp-\bpp\ap -p\,\bu$,
it follows 
\begin{eqnarray}
\fp(\bpp) & = & 
f(\bpp)-p\,\bu\cdot\frac{\partial f(\bpp)}{\partial\bpp} + {\mathcal O}(u^2)
\\ & = &
f(\bpp)\left(1-\frac{\bu \cdot \bp}{p}\frac{\d \ln f}{\d\ln\pp}\right) \,.
\end{eqnarray}
Since $u\equiv|\bu|\ll 1$, the anisotropy induced by the CCG effect is
dominated by the lowest moment, i.e. its dipole moment. Changing to the
differential intensity $I(E)\simeq p^2f(p)$, one obtains 
\begin{equation}
\Ip(\Ep)\simeq I(E)
\left[1+\left(2-\frac{\d\ln I}{\d\ln \Ep}\right)\frac{\bu \cdot \bp}{p}\right].
\end{equation}
Thus the dipole anisotropy due to the CCG effect has the amplitude
\begin{equation}
\dCG\equiv \frac{I_{\rm max}-I_{\rm min}}{I_{\rm max}+I_{\rm
min}}=\left(2-\frac{\d\ln I}{\d\ln E}\right)\,u \,.
\end{equation}
Taking into account the observed spectrum $I(E)\propto E^{-2.7}$ of
cosmic rays above the ankle, $\dCG=(2+2.7)\,u\simeq 0.6\%$. The annual
motion of the  Earth induces only an additional subleading (8\%) modulation 
in the vector $\bu$.

Theoretical predictions of anisotropies for galactic sources depend on
the GMF and the exact source distribution: The amplitude $A$ of galactic
anisotropies increases with energy and may range from 
$A\sim 10^{-4}$ at $E\sim\,{\rm few}\,\times 10^{14}$~eV to 
$A\sim 10^{-2}$ at $E\sim\,{\rm few}\,\times 10^{17}$~eV~\cite{Candia03}.

The flux of extragalactic UHECRs is isotropic in the rest frame
of the CMB at energies $E\lsim E_\ast$
for which the energy-loss horizon $\lambda_{\rm hor}$ of CRs is large 
compared to the scale of inhomogeneities in their source distribution. 
In the same energy range, peculiar velocities average out on cosmological 
scales and the UHECR flux is thus isotropic at leading order.
The exact value of $E_\ast$ depends both on the density of 
the CR sources and on the primary type, but 
$E_\ast\lsim 4\times 10^{19}\,$eV is a conservative estimate: 
For protons $\lambda_{\rm hor}$ is
at the Gpc scale at $E\lsim 10^{19}\,$eV, 
decreasing to about 600~Mpc at $4\times 10^{19}$~eV due to the onset
of the pion production on the CMB, 
and rapidly dropping to few tens of Mpc at larger energies. 
For iron nuclei, $\lambda_{\rm hor}$ abruptly drops below
the Gpc scale only at $E\sim 10^{20}\,$eV when photo-dissociation
processes on the microwave and infrared backgrounds are possible. 
For typical UHECRs source densities of 
$n_s={\rm few}\times 10^{-5}$~Mpc$^{-3}$~\cite{sscf,ns}, 
the number $N_s$ of sources contributing to the
observed flux can be estimated neglecting cosmological effects as
$$
N_s\simeq \frac{4\pi}{3}\lambda_{\rm hor}^3 n_s\simeq 
4\times 10^4 \:\frac{n_s}{10^{-5}\,{\rm Mpc}^{-3}}\,
\left(\frac{\lambda_{\rm hor}}{\rm Gpc}\right)^3 \!\!\!.
$$
Since Poisson fluctuations in $N_s$
are roughly at the 0.5\% level, one might wonder if the CCG effect
could be mimicked by a fluctuation in the number of source per hemisphere.
However, as long as EGMFs wash-out anisotropies,
the dominant intrinsic fluctuation is due to the number of
events $N$ observed at the Earth and not to $N_s$, even for relatively 
low $N_s$. Observational tests that $E_\ast$ was chosen 
low enough are: $i)$ the approximate alignment of the dipole axis of 
$\dCG$ with the one in the CMB; $ii)$ the absence of higher multipole 
moments in the observed maps: While
fluctuations in the number of cosmological sources should lead to 
higher multipole modes $l>1$ with similar intensity $A^{(l)}$, they
are suppressed by powers of $u$ in the case of the CCG effect,
$\dCG^{(l)}\propto u^l$.

The signatures and the properties of the CCG effect are:

{\it (i)}
The amplitude $\dCG$ of the anisotropy is charge- and
energy-independent, as long as the UHECR flux in the energy range
studied is dominated by sources at cosmological distance.

{\it (ii)} 
Since the CCG effect is a dipole anisotropy, the magnitude of
its amplitude should be robust against deflections of UHECRs in the GMF,
and only the dipole axis is displaced. For instance, at energies 2--3$\times
10^{19}$~eV and for proton primaries, the dipole position should be 
aligned to the one observed in the CMB within about 10$^\circ$. 

{\it (iii)}
Observing the CCG feature at only one energy provides combined
information on the intervening GMF and the charge
of the cosmic ray primaries. However, observations at two or more
energies break this degeneracy. For example, the determination of the
average primary charge is straightforward as long as CRs propagate
in the quasi-ballistic regime and given by the ratio of the shifts of 
the CR and CMB diple axis at two different energies.

{\it (iv)}
Moving to lower energies, the anisotropy due to the CCG effect should
disappear as soon as galactic CRs start to dominate. 
Relatively large anisotropies connected to an increased source density
in the disc or towards the galactic center are expected to turn on 
somewhere between $10^{17}$~eV and the ankle~\cite{Candia03}. 
Alternatively, blind regions may induce anisotropies in the
extragalactic CR flux at low energies, as discussed in Sec.~2.

{\it (v)}
Moving to sufficiently high energies,  $\lambda_{\rm hor}$ 
decreases and anisotropies due to local inhomogeneities in the
distribution of sources are expected to dominate. This effect will be
discussed in more detail in the next Section.

Is it possible for present experiments to detect a 0.6\% dipolar
anisotropy in the UHECR flux? In a sample of $N$ events, typical
fluctuations are of the order of $\sqrt{N}$. Thus a 0.6\% level sensitivity
is only reached for $\sqrt{N}/N\simeq 0.006$ or $N\simeq 3\times 10^4$ 
events.
Reference~\cite{Mollerach:2005dv} gave an empirical fit for the 
expected error $\sigma_A$ in the determination of the amplitude of a dipole
anisotropy as function of the event number $N$ and the declination
$\delta$ of the dipole vector, 
$\sigma_A = \sqrt{3/N}\left(1+0.6 \sin^3\delta\right)$,
where a detector located at the PAO site and a maximum zenith angle of 
60$^\circ$ were assumed. This implies that a 3$\,\sigma$ detection
of a 0.6\% anisotropy requires of the order of $10^6$ events and it
thus unlikely for the PAO.
Since the UHECR spectrum is steep, better detection possibilities
are at lower energies, say at 10$^{18}$~eV or below, which will be
explored in the near future by the PAO and especially the TA. 
A negative result would question a transition from galactic
to extragalactic cosmic rays close or below the energy range
considered.

\section{Medium-scale clustering}
\label{msc}

The CCG effect requires that inhomogeneities in
the source distribution of CRs are averaged out. As the free mean
path of CRs decreases for increasing energy, anisotropies connected to
the large-scale structure (LSS) of CR sources become more prominent
and replace the CCG. The exact value of this transition energy
$E_\ast$ depends
both on the amount of clustering in the source distribution and the
free mean path  $\lambda_{\rm CR}$ of CRs, i.e. also the primary type. For
the specific case of proton primaries and a source distribution
proportional to the density of baryons, 
Ref.~\cite{napoli} found $E_\ast\ap 5\times 10^{19}$~eV and
a minimal number of events of order 100 for a detection.

The available data set of UHECR events with published arrival
directions consists of the SUGAR data with energy above $E\geq 1\times
10^{19}$~eV~\cite{SG}, the Yakutsk data as presented at the ICRC
2005~\cite{YK}, the AGASA data set until May 2000 from
Ref.~\cite{AG}, and the HiRes stereo data
set~\cite{Abbasi:2004ib,Abbasi:2004ib_b}.
Additionally,  the arrival directions of six events from Haverah Park,
Volcano Ranch  and Flye's Eye with $E>10^{20}$~eV are given in
Ref.~\cite{reviews}. This data set consists of ${\mathcal O}(100)$ events
and may be already used to test
medium-scale clustering. In this section we review the results of
such a test performed in Ref.~\cite{msc}.

Since the absolute energy scale of each experiment has a rather large
uncertainty, the energies $E$ given by the experiments have to be
shifted to new energies $E'$ to reproduce correctly spectral features
like e.g. the dip. A crucial ingredient of any analysis that combines
data of several experiments and depends on the correct (relative)
assignment of event energies is therefore their consistent rescaling. 

In Ref.~\cite{msc}, the energy rescaling was performed following 
Refs.~\cite{YK,KS03}, but using for convenience the HiRes energies as 
reference energy $E'$.  
The angular two-point auto-correlation function $w$ in
its cumulative version was used as statistical estimator for
possible deviations from an isotropic distribution of arrival
directions. Thus $w$ as function of the angular
scale $\delta$  was defined as    
\be
 w(\delta) = \sum_{i=1}^N\sum_{j=1}^{i-1} \Theta(\delta-\delta_{ij}) \,,
\ee
where $\Theta$ is the step function, $N$ the number of CRs
considered and $\delta_{ij}$ the angular distance between the two
cosmic rays $i$ and $j$. 
Having performed a large sample of Monte Carlo
simulations, the (formal) chance probability $P(\delta)$
to observe a larger value of the autocorrelation
function $w(\delta)$ is the fraction of simulations with $w>w^\ast$, where
$w^\ast$ is the observed value. Since the search and cut criteria were
not fixed a priori, the probabilities  obtained in Ref.~\cite{msc} are
only indicative. But 
they can be used in particular to compare  for different data sets the
relative likelihood to observe the signal as chance fluctuation.

\begin{figure}
\hspace*{-0.4cm}
\epsfig{file=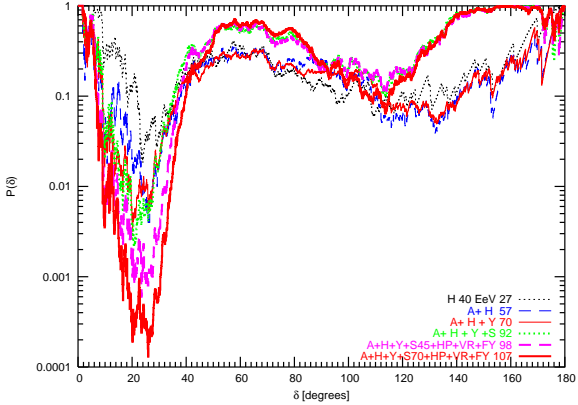,width=0.34\textwidth,angle=270}
\vspace*{-0.6cm}
\caption{
Chance probability $P(\delta)$ to observe a larger value of the autocorrelation
function as function of the angular scale  $\delta$ for different
combinations of experimental data.
\label{p_ch_ex}}
\end{figure}

Figure~\ref{p_ch_ex} shows the chance probability $P(\delta)$ as
function of the angular scale $\delta$ for different combinations of
experimental data. The chance probability $P(\delta)$ shows already a 
$2\sigma$ minimum around 20--30 degrees using only the 27~events of the HiRes
experiments with $E'\geq 4\times 10^{19}$~eV.
Adding more data, the signal around $\delta=25^\circ$
becomes stronger, increasing from $\sim 2\sigma$ for 27~events to $\sim
3.5\sigma$ for 107~events. The position of the
minimum of $P(\delta)$ is quite stable adding more data and every additional 
experimental dataset contributes to the signal. Moreover, 
autocorrelations at scales smaller than $25^\circ$ become more significant  
increasing the dataset. 

To understand better how the search at arbitrary angular scales
influences the significance of the signal, Ref.~\cite{msc} calculated 
the penalty factor for the scan of $P(\delta)$ over
$\delta$. The penalty factor increases for increasing resolution
$\Delta\delta$ of the angular scale $\delta$, but reaches an asymptotic
value for $\Delta\delta\to 0$. The numerical value of the penalty factor
found in the limit $\Delta\delta\to 0$ varies
between 6 for the HiRes data set alone and 30 for the combination
of all data. Since the energy cut was determined by the one chosen 
in Ref.~\cite{Abbasi:2004ib_b}, no additional penalty factor for the
energy had  
to be included. Thus the true probability to
observe a larger autocorrelation signal by chance is $P\approx
3\times 10^{-3}$ for the complete data set.

\begin{figure}
\hspace*{-0.4cm}
\epsfig{file=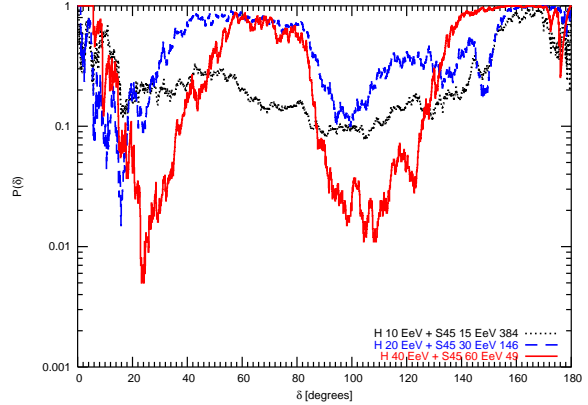,width=0.34\textwidth,angle=270}
\vspace*{-0.6cm}
\caption{
Chance probability $P(\delta)$ to observe a larger value of the autocorrelation
function as function of the angular scale  $\delta$ for different
cuts of the rescaled energy $E'$: black $E'\geq 1\times 10^{19}$~eV,
blue $E'\geq 2\times 10^{19}$~eV and red line $E'\geq 4\times
10^{19}$~eV.
\label{p_ch_E}}
\end{figure}

The addition of data with energy  below $E'\approx 4\times 10^{19}$~eV
reduces the significance of the minimum of $P(\delta)$, 
cf.~Fig.~~\ref{p_ch_E}. For this check one can use only
Hires and SUGAR data, since for the other experiments no arrival
directions for events below $E'=4\times 10^{19}$~eV are
published. Moreover, the energy bin size was dictated by the
one chosen in Ref.~\cite{Abbasi:2004ib_b}.

The reduction of the minimum of $P(\delta)$  could have
various reasons: First, the 
interaction length of the UHECR primaries can increase with decreasing
energy, as in the case of protons or nuclei. Then, the projection on
the two-dimensional skymap averages out more and more three-dimensional
structures. Second, deflections in the Galactic and extragalactic
magnetic fields destroy for lower energies more and more correlations. 
Note that the autocorrelation signal would not disappear lowering the
energy threshold, if it would be caused solely by an incorrect
combination of the exposure of different experiments.

These results, if confirmed by future independent data sets, have
several important consequences\footnote{Reference~\cite{add},
  appearing after the workshop took place, compared the proposed
  signal with the clustering properties expected from the PSCz
  catalogue of galaxies. Its authors argued that the chance
  probability of the signal is consistent within 2 sigma with the
  expectations from the PSCz catalogue. No evidence for a significant
  cross-correlation of the observed events with the LSS overdensities
  was found, which may be explained by deflections in magnetic fields
  and the limited statistics.}. 

Firstly, anisotropies on intermediate angular scales constrain the
chemical composition of UHECRs. Iron nuclei propagate in the 
Galactic magnetic field in a quasi-diffusive regime at $E=4\times
10^{19}$~eV and correlations would be smeared out on scales of 
$\sim 30^\circ$. Therefore, models with a  dominating extragalactic
iron component at the highest energies are disfavored by 
anisotropies on intermediate angular scales. Since the relative
deflections of CRs in the same energy range is reduced, a conclusive
statement requires however a larger event sample.

Secondly, the probability that small-scale clusters are indeed from
point sources will be reduced if the clusters are in regions with an
higher UHECR flux. By contrast, the 
observation of clusters in the "voids"  would be less
likely by chance than in the case of an UHECR flux without medium
scale anisotropies. 

However, the most important consequence of these findings is the prediction
that astronomy with UHECRs is possible at the highest energies. The
minimal energy required seems to be around $E'=4\times 10^{19}$~eV, 
because at lower energies UHECR arrive more and more isotropically. 
This trend is expected, because at lower energies both deflections 
in magnetic fields and the average distance $l$ from which UHECRs can
arrive increase. 
Since the two-dimensional skymap corresponds to averaging all 
three-dimensional structures (with typical scale $L$) over the distance
$l$, no anisotropies apart from the CCG effect are expected for $l\gg
L$. Thus, if the signal
found in Ref.~\cite{msc} will be confirmed it has to be related to the local
large scale structure.

\section{Small-scale clustering}
\label{ssc}

The presence of small-scale cluster, i.e.\ cluster of events within
the angular resolution of UHECR experiments, was first noticed for the
AGASA data in Ref.~\cite{ssca}. About 20\% of the world data set
with energy $\geq 4\times 10^{19}$~eV measured at that time were
clustered in angular doublets or even triplets; both triplets are
found near the  supergalactic plane. The chance probability to observe the
clustered events in the case of an isotropic distribution of arrival
directions was estimated to be $<1\%$~\cite{sscb}.

In particular the AGASA data set consisting of four pairs and one
triplet within $2.5^\circ$ out of $N=57$ CRs was analyzed by various
groups~\cite{ssca,sscb,ssccd,ssce,sscf}. The statistical significance 
ascribed to the clustering signal varies however strongly in different
analyzes~\cite{ssca,sscb,ssccd,ssce,sscf}. 
This variance is explained partly by the
difficulty of assigning a posteriori a chance probability to a search
for a signal without prescribing a priori the cuts. Moreover, the
HiRes experiment~\cite{Abbasi:2004ib} has not confirmed clustering
yet, but this finding is still compatible with
expectations~\cite{ssce,sscf}. The 
preliminary data of the PAO have been searched only
for single sources, with negative result~\cite{Revenu:2005}.
From our discussion of the influence of the GMF on the arrival
direction of UHECRs it is clear that the expected deflection towards
the GC are larger. Moreover, they depend strongly on the size of the
dipole field. It may be therefore essential for searches of point
sources using the data of the PAO to apply models of the GMF and to
correct for the resulting deflections.

Finally, there is the open question of correlations of UHECR arrival
directions with BL Lacs, put forward first in Ref.~\cite{corrb}.
Additionally to the first claim of proton primaries correlated with
the AGASA and Yakutsk data, Ref.~\cite{corrd} found a correlation of a
neutral component at lower energies with the HiRes data. The status of
these correlations is still disputed at present; a summary of the
various correlation claims and also a forecast can be 
found in Ref.~\cite{corre}.

\section{Summary and conclusions}

Possible anisotropies expected for extragalactic cosmic rays can be
classified into four subclasses: 
$i)$ At such high energies that
deflections in extragalactic magnetic fields are sufficiently small 
point sources may reveal
themselves as small-scale clusters of UHECR arrival directions. This
requires additionally a rather low density of UHECR sources so that
the probability to observe several events of at least a subset of
especially bright sources is large enough.
$ii)$ Moving to lower energies, the energy-loss horizon of UHECRs and
thereby the number of sources visible increases. Moreover, deflections
in magnetic fields become more important. As a result, the
identification of single sources is not possible anymore. Instead,
anisotropies on medium scales should reflect the inhomogeneous
distribution of UHECR sources that is induced by the observed LSS 
structure of matter.
$iii)$ At even lower energies, also the LSS structure of
sources disappears, both because the inhomogeneities in the source
distribution will be averaged out due to the increased  energy-loss
horizon of UHECRs  and because of deflections in the EGMF. Thus the CR
sky would 
appear isotropic, if the Earth would be at rest with respect to the
cosmological rest frame. As the observation of the CMB dipole shows,
this is not case, and a dipole anisotropy of 0.6\% is expected if the
CR flux is dominated by sources at cosmological distance.
The shift of the dipole as function of energy provides information
about the mean charge of CRs and the GMF.
$iv)$ Finally, the GMF can induce anisotropies in the observed flux of
extragalactic UEHCRs (even if it is isotropic at the boundary of the
Milkyway), for rigidities low enough that blind regions exist. 
According to the estimates of Ref.~\cite{kst} anisotropies of this kind
should be expected in models that invoke a dominating extragalactic
proton component already at $E\simeq 4\times 10^{17}$~eV  or
extragalactic iron nuclei at $E\lsim 10^{19}$~eV.

It is not guaranteed that all these four anisotropies can be observed. 
If EGMFs are large, UHECR primaries are nuclei and/or the source density
is large, the integrated flux above the energy where point sources
become visible may be for the present generation of UHECR experiments 
too small. Similarly, a transition from galactic to extragalactic sources 
at a relatively high energy reduces the chances to observe the CCG
effect. Experimentally, the easiest accessible anisotropy
is the medium-scale anisotropy connected to the LSS of UHECR sources.
Here, the event number required to confirm the proposed signal in
Ref.~\cite{msc} should be collected within one or two years by the PAO.

\section*{Acknowledgments}
It is a pleasure to thank Dima Semikoz and Pasquale Serpico for
fruitful collaborations on which this talk is based on, and
to thank Dima Semikoz also for comments on this draft.


\end{document}